\documentclass[journal,comsoc]{IEEEtran}
\usepackage{amsmath,amsfonts,amssymb}
\usepackage{cite}
\usepackage[pdftex]{graphicx}
\usepackage{color}
\usepackage{cases}
\usepackage[T1]{fontenc}
\usepackage{algorithm}
\usepackage{algpseudocode}
\allowdisplaybreaks 
\newcommand \FigWidth {1.0 \columnwidth}

\begin{document}
\title{Secure Power Control for Downlink Cell-Free Massive MIMO With Passive Eavesdroppers}

\author{Junguk~Park,~\IEEEmembership{Student Member,~IEEE,}
	Sangseok Yun,~\IEEEmembership{Member,~IEEE,}
        and~Jeongseok~Ha,~\IEEEmembership{Senior Member,~IEEE} 

\thanks{ J. Park and J. Ha are with the School of Electrical Engineering, Korea Advanced Institute of Science and Technology, Daejeon 34141, South Korea (e-mail: pjuk@kaist.ac.kr; jsha@kaist.edu). 

S. Yun is with the Department of Information and Communications Engineering, Pukyong National University, Busan 48513, South Korea, e-mail: ssyun@pknu.ac.kr. } } 
 
\maketitle
\begin{abstract}
This work studies secure communications for a cell-free massive multiple-input multiple-output (CF-mMIMO) network which is attacked by multiple passive eavesdroppers overhearing communications between access points (APs) and users in the network. It will be revealed that the distributed APs in CF-mMIMO allows not only legitimate users but also eavesdroppers to reap the diversity gain, which seriously degrades secrecy performance. Motivated by this, this work proposes an artificial noise (AN)-aided secure power control scheme for CF-mMIMO under passive eavesdropping aiming to achieve a higher secrecy rate and/or guarantee security. In particular, it will be demonstrated that a careful use of AN signal in the power control is especially important to improve the secrecy performance. The performance of the proposed power control scheme is evaluated and compared with various power control schemes via numerical experiments, which clearly shows that the proposed power control scheme outperforms all the competing schemes. 
\end{abstract}


\section{Introduction}

Recently, cell-free massive multiple-input multiple-output (CF-mMIMO) has been gaining attention as a promising network structure for beyond 5G (B5G) networks, since it can provide uniformly high quality of service for all users throughout the coverage area \cite{Ngo15Cell-Free, Zhang19Cell-free}. In contrast to conventional \emph{co-located} massive MIMO (mMIMO) which co-locates a massive number of antennas at an access point (AP), CF-mMIMO has a massive number of APs distributed across a network. In addition, all APs in the network cooperatively serve all users. These features of CF-mMIMO provide the spatial diversity gain and the array gain resulting from the distributed APs and the cooperative beamforming, respectively \cite{Ngo17Cell-free}.

The broadcast nature of wireless channels makes challenges in securing wireless communication systems. Physical layer security (PLS) has been considered as one of enabling technologies for constructing secure communication systems in B5G networks \cite{Nguyen21Security}. Unlike conventional security measures based on computational cryptography, PLS utilizes physical layer resources to achieve security goals. In addition, PLS builds on the information-theoretic security which brings inherent advantages, such as unconditionally perfect secrecy and scalability. Thus, there have been notable efforts to construct PLS-based secure CF-mMIMO systems, such as secure power control schemes \cite{Timilsina18Physical, Hoang18Cell-free, Zhang20Secure, Zhang22Secure} and pilot spoofing attack detection \cite{Zhang19Secrecy}.

In \cite{Timilsina18Physical}, the achievable secrecy rate in the presence of an active eavesdropper was analyzed. In particular, the authors proposed a power control scheme with artificial noise (AN) for maximizing the achievable secrecy rate of the user under attack. Meanwhile, the authors in \cite{Hoang18Cell-free} considered two different power control problems without AN in the presence of an active eavesdropper. They formulated optimization problems to maximize secrecy rate of the attacked user and minimize the power consumption. In \cite{Zhang20Secure}, the authors investigated the impact of hardware impairments on an achievable secrecy rate under pilot spoofing attack. It was also investigated the impact of Rician fading and low-resolution digital-to-analog converters (DACs) on the secrecy performance under pilot spoofing attack in \cite{Zhang22Secure}. Secrecy performance in simultaneous wireless information and power transfer (SWIPT) CF-mMIMO with an active eavesdropper has been studied in \cite{Alageli20Optimal}. Recently, reconfigurable intelligent surface (RIS) assisted secure CF-mMIMO was investigated in \cite{Elhoushy22Exploiting}.

The aforementioned studies of PLS for CF-mMIMO focus on an \emph{active} eavesdropping attack called pilot spoofing/contamination attack \cite{Zhou12Pilot} in which an active eavesdropper transmits the same pilot signal as the one for a target user to tilt the direction of the beamforming towards the attacker. Contrary to the active eavesdropper, a passive eavesdropper does not tamper with the signal of the target users, e.g., pilot signals, and instead passively overhears the transmitted signals to the target user. It is well known \cite{Kapetanovic15Physical} that  co-located mMIMO selectively provides the benefit of array gain to legitimate users, which allows the legitimate users to achieve secrecy capacity close to the capacity of ordinary communication without further effort for secrecy, e.g., AN signals \cite{Kapetanovic15Physical}. 

It seems that a large number of transmit antennas in CF-mMIMO, while distributed across a network, also empower the legitimate parties to nullify the passive eavesdropping as observed in co-located mMIMO. This is in part why the passive eavesdropping in CF-mMIMO has not been touched yet. However, contrary to co-located mMIMO, as the number of APs in CF-mMIMO increases, not only legitimate users but also eavesdroppers have the benefit of the spatial diversity, which exposes legitimate users in CF-mMIMO to the threat of passive eavesdropping. This motivates us to investigate a novel secure power control scheme for CF-mMIMO under the passive eavesdropping. In particular, we will demonstrate the vulnerability of CF-mMIMO to the passive eavesdropping and also develop an AN-aided secure power control scheme for CF-mMIMO attacked by multiple passive eavesdroppers. To formulate the secure power control as an optimization problem, we derive a lower bound on the secrecy rate with AN in a closed-form. Then, we devise the AN-aided secure power control algorithm which maximizes the minimum secrecy rate among all users. For comprehensive performance comparisons, we consider an existing power control scheme in \cite{Ngo17Cell-free} which maximizes the minimum data rate among all users. The comparisons reveal that the power control scheme in \cite{Ngo17Cell-free} can not guarantee the secure fairness, i.e., the minimum secrecy rate among all users is zero. The impact of AN on the minimum secrecy rate is also analyzed in the comparisons, which enlightens when the AN signal is indispensable.

\section{System Model}
We consider a downlink CF-mMIMO system in which there are {\it{M}} single antenna APs, {\it{K}} single antenna users, and {\it{J}} single antenna eavesdroppers (Eves). The block-fading channel model and a time-division duplex (TDD) operation with channel reciprocity are assumed. The TDD coherence block consisting of $\tau_c$ symbols, is divided into three parts: the uplink channel estimation, the uplink data transmission, and the downlink data transmission. The length of the three parts are denoted by $\tau_p$, $\tau_u$, and $\tau_d$ symbols, respectively. For focusing on the downlink CF-mMIMO, in this work, the uplink data transmission is not considered, that is, $\tau_u=0$. All Eves are assumed to be a passive eavesdropper and not to collude with each other. Thus, all Eves do not transmit any signals, and only overhear the transmitted signals at the downlink data transmission phase.

The channel coefficients from the $m$th AP to the $k$th user and from the $m$th AP to the $j$th Eve are denoted by $g_{mk}$ and $g_{mj}^e$, respectively. Then, the channel coefficients $g_{mk}$ and $g_{mj}^e$ are given by $g_{mk}=\sqrt{\beta_{mk}}h_{mk}$ and $g_{mj}^e=\sqrt{\beta_{mj}^e}h_{mj}^e$, 
where $\beta_{mk}$ and $\beta_{mj}^e$ represent the large-scale fading, and $h_{mk}$ and $h_{mj}^e$ represent the small-scale fading, respectively. In this work, we assume independent and identically distributed (i.i.d.) Rayleigh fading channel. That is, the channel gains, $h_{mk}$ and $h_{mj}^e$  follow zero mean unit variance complex Gaussian distribution, $\mathcal{CN}(0,1)$.

In the uplink channel estimation phase, all users simultaneously transmit their own pilot signals. All pilot signals are assumed to be orthonormal each other, and then the received signal at the $m$th AP can be represented as 
\[
	\mathbf{y}_{p,m} = \sqrt{\tau_p p_p} \sum_{k = 1}^K g_{mk}\boldsymbol{\psi}_k + \mathbf{w}_{p,m},
\]
where $p_p$, $\boldsymbol{\psi}_k  \in \mathbb{C}^{1 \times \tau_p} $, and $\mathbf{w}_{p,m} \in \mathbb{C}^{1 \times \tau_p}$ respectively represent the pilot transmit power, the pilot signal of the $k$th user, and the thermal noise vector whose elements follow i.i.d. $\mathcal{CN} (0,1)$. Then, each AP separately estimates the channel between itself and each user based on the linear minimum mean square error (LMMSE) estimation. The LMMSE estimate of $g_{mk}$ is obtained by
\begin{equation}
 \hat{g}_{mk}  =  \mathbb{E}\left[ g_{mk}y_{p,mk} \right] \big(\mathbb{E}\left[ | y_{p,mk}|^2 \right]\big)^{-1} y_{p,mk} = c_{mk} y_{p,mk}, \label{Eq:g_mk}
\end{equation}
where
\[
	y_{p,mk}  = \mathbf{y}_{p,m} \boldsymbol{\psi}^{H}_k \text{ and }  c_{mk} = \frac{\sqrt{\tau_p p_p} \beta_{mk}}{\tau_p p_p \sum_{k '=1}^K \beta_{mk'} + 1}. 
\]

In the downlink data transmission phase, each AP transmits the data signals for all users. It is assumed that each AP uses the conjugate beamforming for the data signals \cite{Ngo17Cell-free} and the random beamforming for AN signal \cite{Timilsina18Physical,Zhang22Secure}. The transmitted signal from the $m$th AP can be expressed as 
\[
	x_m = \sum_{k=1}^K \sqrt{p_{mk}} \hat{g}^*_{mk} s_k + \sqrt{p_{mv}} v_{m}, 
\]
where $p_{mk}$ and $p_{mv}$ are the transmit powers of data for the $k$th user and AN at the $m$th AP, respectively.
The data signal for the $k$th user and the AN signal at the $m$th AP are denoted by $s_k \sim \mathcal{CN} (0,1)$ and $v_m \sim \mathcal{CN} (0,1)$, respectively. The total transmit power of the data and AN signals is limited in such a way that $\mathbb{E}\left[ \left| x_m \right|^2 \right] \le p_t$ for $m \in \{1, 2, ..., M\}$.

The received signals at the $k$th user and the $j$th Eve are respectively given by
\begin{align}
  & r_k = \sum_{m=1}^M g_{mk}x_m+w_{d,k} =  \sum_{\substack{k'=1}}^K f_{kk'} s_{k'} + z_{k}+ w_{d,k}, \label{Eq:r_k} \\
  & r_j^e = \sum_{m=1}^M g_{mj}^e x_m+w_{d,j}^e = \sum_{\substack{k=1 }}^K f_{jk}^e s_{k} + z_{j}^e+ w_{d,j}^e, \label{Eq:r_e_j}
\end{align}
where $f_{kk'} =  \sum_{m=1}^M \sqrt{p_{mk'}}\hat{g}^*_{mk'} g_{mk}$, $z_{k} = \sum_{m=1}^M \sqrt{p_{mv}} g_{mk} v_m$, $f_{jk}^e = \sum_{m=1}^M \sqrt{p_{mk}}\hat{g}^*_{mk} g_{mj}^e$, $z_{j}^e = \sum_{m=1}^M \sqrt{p_{mv}} g_{mj}^e v_m$, and $w_{d,k}$ and $w_{d,j}^e$ are the thermal noises following $\mathcal{CN} (0,1)$. 

\section{Artificial Noise-Aided Secure Power Control} \label{Sec:ANSPC}
In this section, we formulate an optimization problem which finds the optimal power control coefficients for the data and AN at each AP. In the formulation of the problem, we need an analytic expression of the secrecy rate, which however, to the best of our knowledge, is impossible to obtain. Thus, we instead derive a lower bound on the secrecy rate with a lower and upper bounds on user and leakage rates, respectively. For the rate of user, $R_k$, we first apply the use-and-then-forget (UatF) lower bound \cite{Bjornson17Massive} to $R_k$, which results in a lower bound on $R_k$, denoted by $\bar{R}_k$ in \eqref{Eq:R_k}. Meanwhile, for the leakage rate, $R_{jk}^{e}$, we derive an upper bound, denoted by $\bar{R}_{jk}^e$ by assuming the worst-case scenario that Eves perfectly know their channel gains \cite{Timilsina18Physical, Hoang18Cell-free, Zhang20Secure, Zhang22Secure}. Then, we have the upper bound, $\bar{R}_{jk}^e$ in \eqref{Eq:R_eve}.
\begin{align}
	& R_k \ge \bar{R}_k = \log_{2} \left(1 +  \frac{DS_k}{BU_k + \sum_{\substack{k'=1 \\ k' \neq k}}^{K} UI_{kk'} + AN_k + 1} \right), \label{Eq:R_k} \\
	& R_{jk}^{e}\le \bar{R}_{jk}^e =  \log_{2} \left(1 +  \frac{LS_{jk}}{\sum_{\substack{k'=1 \\ k' \neq k}}^{K} LS_{jk'} + AN_{j}^e + 1} \right), \label{Eq:R_eve}
\end{align}
where $DS_k$, $BU_k$, $UI_{kk'}$, $AN_k$, $LS_{jk}$, and $AN_{j}^e$ are shown at the top of the next page. 

\begin{figure*}[ht]
\begin{align}
	& DS_k = \left( \sum_{m=1}^M \sqrt{p_{mk}} \gamma_{mk} \right)^2, \;
	BU_k =  \sum_{m=1}^M p_{mk} \gamma_{mk} \beta_{mk}, \;
	UI_{kk'} = \sum_{m=1}^M p_{mk'} \gamma_{mk'} \beta_{mk}, \;
	AN_k =  \sum_{m=1}^M p_{mv} \beta_{mk}, \; 
	LS_{jk} = \sum_{m=1}^M p_{mk} \gamma_{mk} \beta_{mj}^e, \; \nonumber \\
	& AN_{j}^e = \sum_{m=1}^M p_{mv} \beta_{mj}^e, \;
	SINR_k = \frac{DS_k}{IN_k} = \frac{DS_k}{BU_k + \sum_{\substack{k'=1 \\ k' \neq k}}^{K} UI_{kk'} + AN_k + 1}, \;
	SINR_{jk}^e = \frac{LS_{jk}}{IN_{jk}^e} =  \frac{LS_{jk}}{\sum_{\substack{k'=1 \\ k' \neq k}}^{K} LS_{jk'} + AN_{j}^e + 1}. \; \label{Eq:Rate_terms}
\end{align}
\hrulefill
\end{figure*}

The detailed derivations are provided in Appendix \ref{Sec:App_rate}. Finally, we obtain a lower bound on the secrecy rate as follows:
\[
	R_{jk}^{s} = \left[ \bar{R}_k - \bar{R}_{jk}^e \right]^{+},\, \mathrm{for} \, k \in \{1,2,...,K\} \text{ and } j \in \{1,2,...,J\},
\]
where $\left[ x \right]^+$ is $\mathrm{max}\left[ 0, x \right]$.

The optimization problem for the secure fairness, i.e., maximizing the minimum secrecy rate of all users, can be formulated as
\begin{subequations}\label{Eq:opt_1}
\begin{align} 
	\max_{\substack{\{p_{mk}\},\{p_{mv}\}}} & \min_{\substack{j, k}}  \bar{R}_k - \bar{R}_{jk}^e \label{Eq:opt_obejct} \\
	 \text{s.t. } & \textstyle \sum_{k=1}^{K} p_{mk}\gamma_{mk} + p_{mv} \le p_t, \; \forall m, \label{Eq:power_const} \\
	& p_{mk}\ge 0, \; \forall k, \; \forall m, \label{Eq:p_mk_const} \\
	& p_{mv}\ge 0, \; \forall m, \label{Eq:p_mv_const}
\end{align}
\end{subequations}
where the constraints in \eqref{Eq:power_const} comes from $\mathbb{E}\left[ \left| x_m \right|^2 \right] \le p_t$ for $m \in \{1,2,...,M\}$. By introducing slack variables $t$ and $\{\omega_{jk}\}$, we reformulate the optimization problem in \eqref{Eq:opt_1} as follows:
\begin{subequations}\label{Eq:opt_2}
\begin{align} 
	\max_{\substack{\{p_{mk}\}, \{p_{mv}\}},\{\omega_{jk}\}, t}& \;  t  \label{Eq:opt_2_obejct} \\
	\text{s.t. } & \bar{R}_k  -  \omega_{jk} \ge t,  \; \forall k,  \; \forall j, \label{Eq:opt_2_user_rate_const} \qquad\quad \\
	&  \bar{R}_{jk}^e \le \omega_{jk},   \; \forall k, \; \forall j, \label{Eq:opt_2_Eve_rate_const} \\
	&  \eqref{Eq:power_const}, \; \eqref{Eq:p_mk_const}, \; \eqref{Eq:p_mv_const}.
\end{align}
\end{subequations}

Unfortunately, the constraints in \eqref{Eq:opt_2_user_rate_const} and \eqref{Eq:opt_2_Eve_rate_const} are non-convex. Thus, we will solve the optimization problem in \eqref{Eq:opt_2} with an iterative method solving a sequential convex problems. To resolve the non-convexity, we replace $\bar{R}_k$ and $\bar{R}_{jk}^e$ with their linear approximations at each iteration. In particular, the constraints in \eqref{Eq:opt_2_user_rate_const} and \eqref{Eq:opt_2_Eve_rate_const} at the $(n+1)$th iteration can be approximated as 
\begin{subequations}
\begin{align}
	& \bar{R}_k^{[n]} + \frac{1}{\ln2} \frac{SINR_k^{[n]}}{1+SINR_k^{[n]}}\left( 2\sqrt{\frac{DS_k}{DS_k^{[n]}}} - \frac{IN_k}{IN_k^{[n]}}-1 \right)   \ge t + \omega_{jk}, \label{Eq:opt_user_rate_const_lower} \\
	& \bar{R}_{jk}^{e,[n]}+ \frac{1}{\ln2}\frac{SINR_{jk}^{e,[n]}}{1+SINR_{jk}^{e,[n]}}\left( \frac{LS_{jk}}{LS_{jk}^{[n]}} - \frac{IN_{jk}^{e}}{IN_{jk}^{e,[n]}} \right)  \le \omega_{jk}, \label{Eq:opt_Eve_rate_const_app}
\end{align}
\end{subequations}
where $\bar{R}_k^{[n]}$, $\bar{R}_{jk}^{e,[n]}$, $SINR_k^{[n]}$, $DS_k^{[n]}$, $IN_k^{[n]}$, $SINR_{jk}^{e,[n]}$, $LS_{jk}^{[n]}$, and $IN_{jk}^{e,[n]}$ can be obtained by respectively replacing $\{p_{mk}\}$ and $\{p_{mv}\}$ in \eqref{Eq:R_k}, \eqref{Eq:R_eve}, and \eqref{Eq:Rate_terms} with $\{p_{mk}^{[n]}\}$ and $\{p_{mv}^{[n]}\}$ which are solutions at the $n$th iteration. The detailed derivations are provided in Appendix \ref{Sec:App_bound}. Then, \eqref{Eq:opt_user_rate_const_lower} and \eqref{Eq:opt_Eve_rate_const_app} become convex constraints at each iteration. Finally, the optimization problem in \eqref{Eq:opt_2} can be reformulated as 
\begin{subequations}\label{Eq:opt_3}
\begin{align} 
	& \max_{\substack{\{p_{mk}\}, \{p_{mv}\}},\{\omega_{jk}\}, t} \;  t  \label{Eq:opt_3_obejct} \\
	& \text{s.t. } \eqref{Eq:power_const}, \; \eqref{Eq:p_mk_const}, \; \eqref{Eq:p_mv_const}, \; \eqref{Eq:opt_user_rate_const_lower}, \; \eqref{Eq:opt_Eve_rate_const_app}.  \label{Eq:opt_3_constraints}
\end{align}
\end{subequations}
The reformulated optimization problem in \eqref{Eq:opt_3} is convex and can be solved by utilizing the convex optimization tools, e.g., CVX. At each iteration, we update the linear approximations, i.e., \eqref{Eq:opt_user_rate_const_lower} and \eqref{Eq:opt_Eve_rate_const_app}, based on the solutions at the previous iteration, until $t$ converges to a fixed value, i.e., $|t^{[n]}-t^{[n-1]}| \le \epsilon$ where $\epsilon$ is a convergence constraint.  The proposed algorithm guarantees to converge since $t^{[n+1] }\ge t^{[n]}$ will always hold and the maximum value of $t$ is bounded by the transmit power constraint in \eqref{Eq:power_const}. The proposed algorithm is outlined in Algorithm \ref{Alg:Outline}. 

\begin{algorithm}[!t]
\caption{The AN-aided secure power control algorithm} \label{Alg:Outline}
    \begin{algorithmic}[1]
        \State Initialize $n \gets 0$, $t^{[0]} \gets 0$, $\{p_{mk}^{[0]}\}$, $\{p_{mv}^{[0]}\}$, 
        \Repeat
            \State $n \gets n+1$
            \State Solve the optimization problem in \eqref{Eq:opt_3}, and obtain the solutions of $\{p_{mk}^{*}\}$, $\{p_{mv}^{*}\}$, and $t^*$.
            \State Update $\{p_{mk}^{[n]}\} \gets \{p_{mk}^{*}\}$, $\{p_{mv}^{[n]}\} \gets \{p_{mv}^{*}\}$, $\{t^{[n]}\} \gets \{t^{*}\}$
	    \State Update the initial points in \eqref{Eq:opt_user_rate_const_lower} and \eqref{Eq:opt_Eve_rate_const_app}
        \Until $|t^{[n]}-t^{[n-1]}| \le \epsilon$.
    \end{algorithmic}
\end{algorithm}

\vspace{-10mm}
\begin{table}[!b] 
\caption{Simulation parameters}
\label{Tbl:Simulation_para}
\centering
\begin{tabular} {|c|c|}
\hline
\bf{Parameter} & \bf{Value} \\ \hline
Carrier frequency & 1.9 GHz \\ 
Reference distances for path-loss & 10 m, 50 m \\
AP and UE heights  & 15 m, 1.65 m \\
TDD parameters per samples &  $\tau_c=200$, $\tau_p=K$, $\tau_d= 200-K$ \\ 
Transmit powers &  $p_p=200$ mW, $p_t=100$ mW \\
Thermal noise power & -94 dBm \\ 
Convergence constraint & $\epsilon=0.01$ \\ \hline
\end{tabular}
\end{table}

\begin{figure*}
\centering
 \begin{minipage}[t]{.315\linewidth}
  \includegraphics[width=\FigWidth]{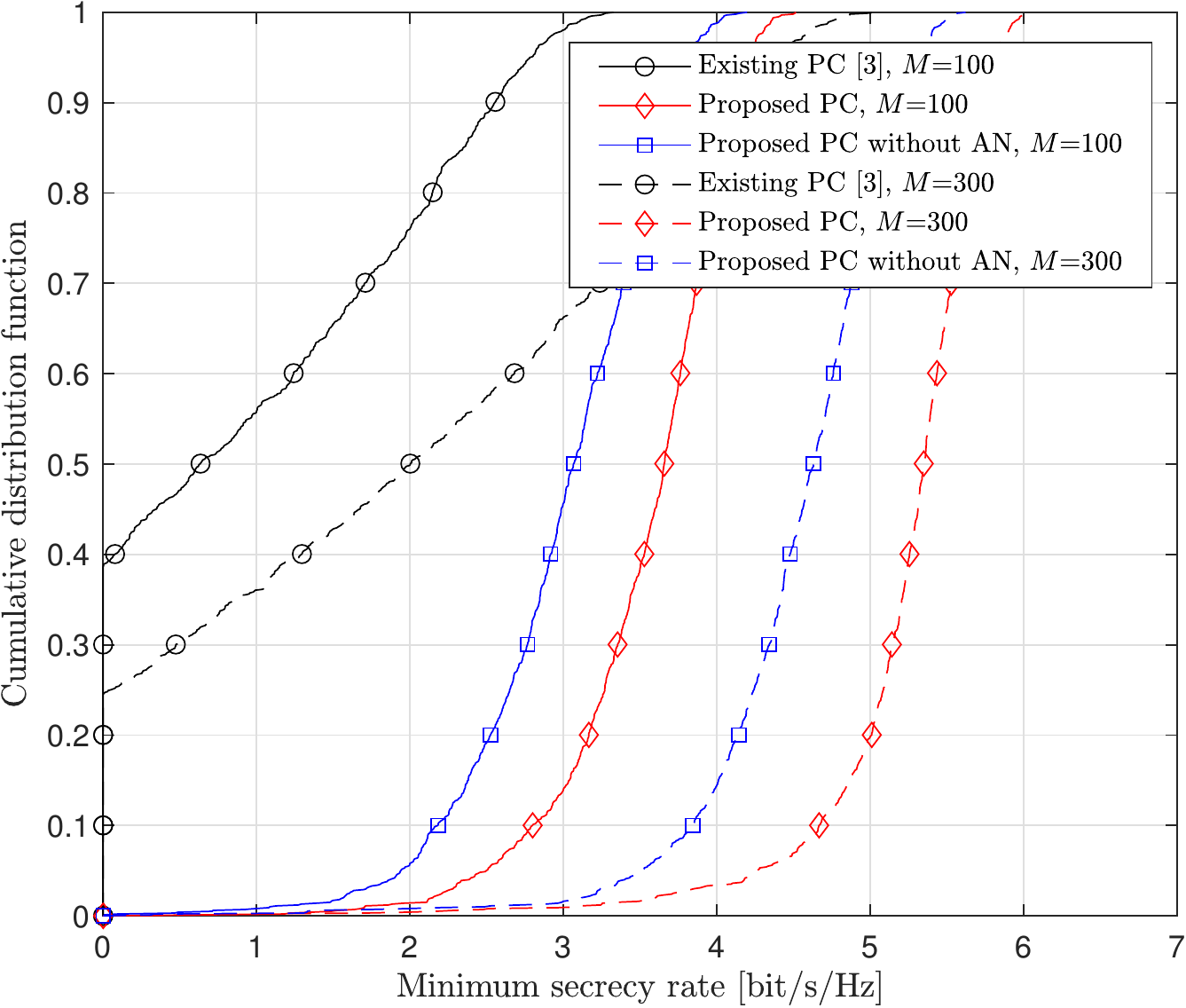}
  \caption{The CDF of the minimum secrecy rate when $J=1$ and $K=2$.} \label{Fig:Min_sec_cha_M}
  \end{minipage}\hfil
\begin{minipage}[t]{0.315\linewidth}
  \includegraphics[width=\FigWidth]{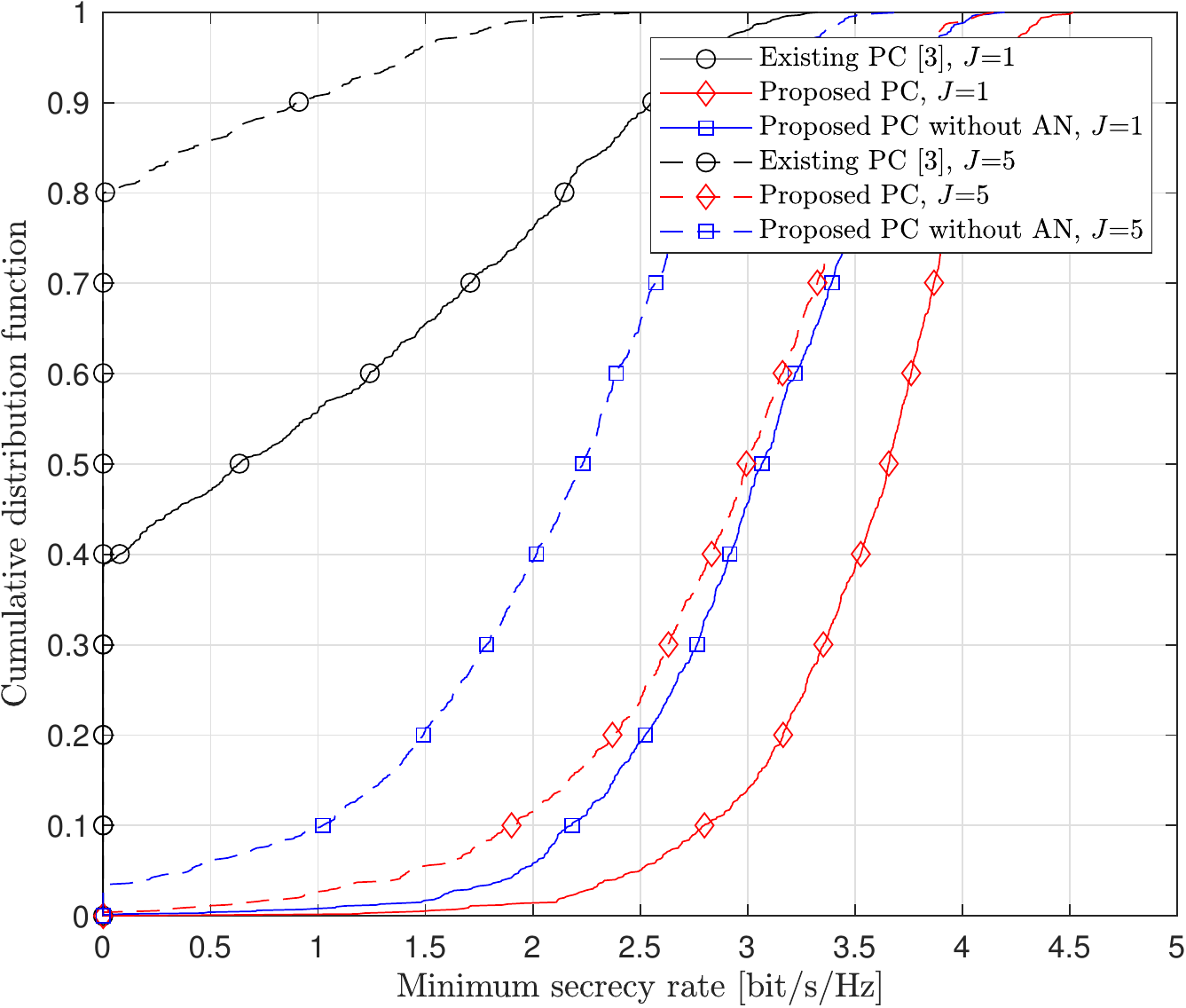}
  \caption{The CDF of the minimum secrecy rate when $M=100$ and $K=2$.} \label{Fig:Min_sec_cha_J}
\end{minipage}\hfil
\begin{minipage}[t]{0.315\linewidth}
  \includegraphics[width=\FigWidth]{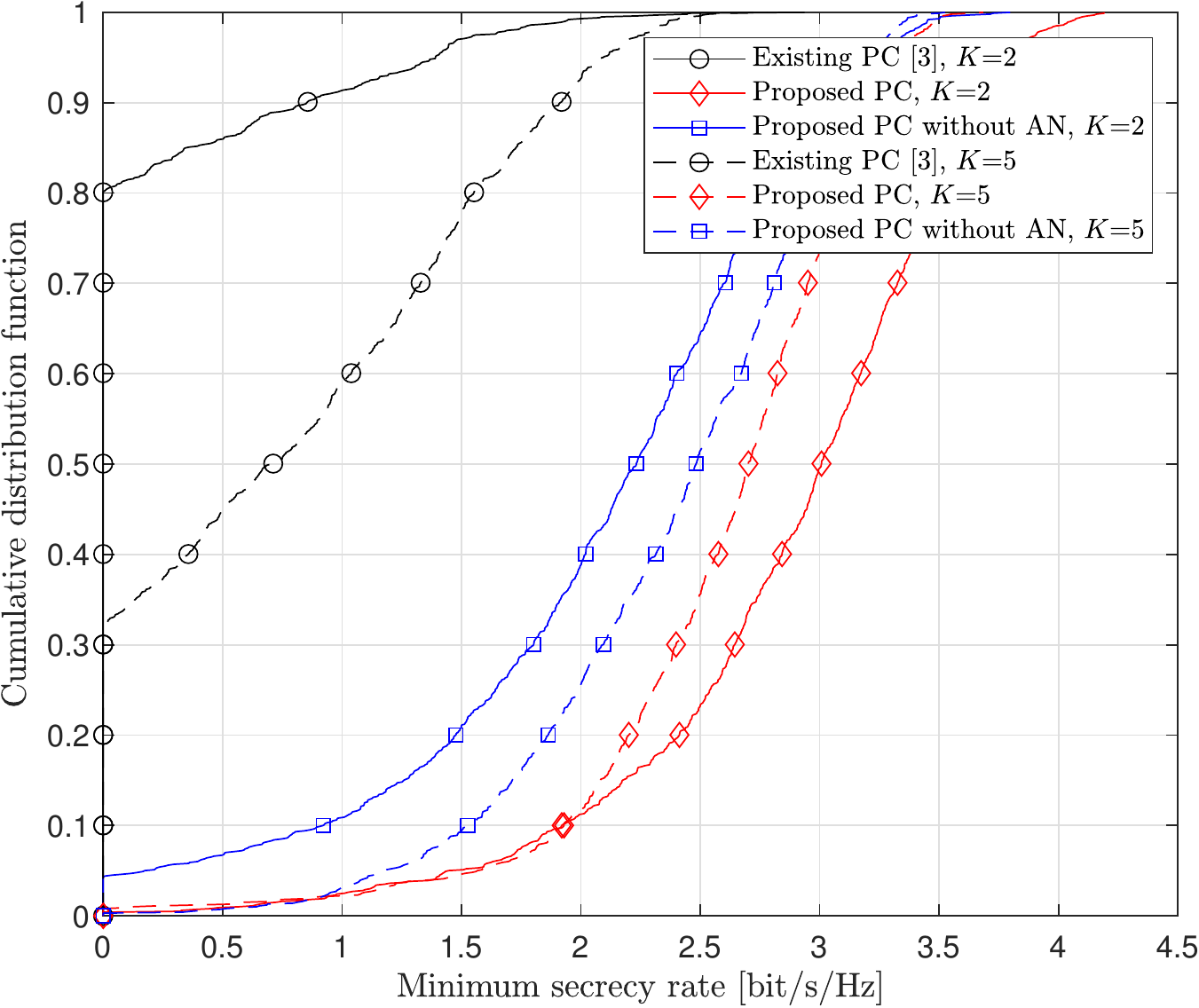}
  \caption{The CDF of the minimum secrecy rate when $M=100$ and $J=5$.} \label{Fig:Min_sec_cha_K}
\end{minipage}
\end{figure*}

\section{Simulation Results} \label{Sec:Sim}

In this section, we carry out  Monte-Carlo simulations to evaluate and analyze the secrecy performance of the proposed AN-aided power control scheme. For the simulations, we consider a network in which $M$ APs,  $K$ users, and $J$ Eves are randomly distributed in a square area of 1 $\times$ 1 $\mathrm{Km^2}$. 
For the propagation and path-loss model, the COST-Hata model and the three-slope path-loss model are used \cite{Ngo17Cell-free}. We carry out simulations for 1000 random realizations of the network deployment. The detailed simulation parameters are summarized in Table \ref{Tbl:Simulation_para}.

The initial values of $p_{mk}$ and $p_{mv}$, denoted by $p_{mk}^{[0]}$ and $p_{mv}^{[0]}$, respectively, are required for the initialization step in Algorithm \ref{Alg:Outline}. We extend the heuristic power control scheme in \cite{Interdonato19Scalability} such that it can simultaneously transmit the AN signal for the purpose of security, and adopt them as the initial values as follows:
\begin{align} 
p_{mk}^{[0]} &= \frac{p_t}{\sqrt{\gamma_{mk}} \left( \sum_{k=1}^{K}\sqrt{\gamma_{mk}} + \sum_{j=1}^{J}{\sqrt{\beta_{mj}^{e}}}\right)}, \label{Eq:p_mk_initial} \\ 
p_{mv}^{[0]} &= \frac{p_t \sum_{j=1}^{J}\sqrt{\beta_{mj}^{e}}}{\sum_{k=1}^{K}\sqrt{\gamma_{mk}} + \sum_{j=1}^{J}\sqrt{\beta_{mj}^{e}}}.  
\label{Eq:p_mv_initial} 
\end{align}
The heuristic secure power control scheme in \eqref{Eq:p_mk_initial} and \eqref{Eq:p_mv_initial} implies that a more transmit power is allocated to the data (AN, resp.) signal when users (Eves, resp.) are closely located to the AP.

For the comprehensive performance comparison, we consider two competing power control schemes: 1) the existing power control scheme in \cite{Ngo17Cell-free} which only maximizes the minimum data rate among all users without considering security, 2) the proposed secure power control scheme without AN which only optimizes ${p_{mk}}$ at $p_{mv}=0$ for $m\in \{1,2...,M\}$ by using our proposed power control algorithm.

The result in Fig. \ref{Fig:Min_sec_cha_M} shows the comparisons between the proposed AN-aided secure power control scheme and the two competing power control schemes. The comparisons are conducted from the perspective of the cumulative distribution function (CDF) of the minimum secrecy rate among all users at different values of $M$. One can find that the proposed AN-aided secure power control scheme always shows the best secrecy performance. It is also observed that the existing scheme \cite{Ngo17Cell-free} can not guarantee the secure fairness, i.e., the user with the minimum secrecy rate is zero. Specifically, the existing scheme suffers from the secure fairness outage with a probability of 24.52\% even for the case of $M=300$ whereas the proposed scheme with AN always provides non-zero minimum secrecy rate. The results in Fig. \ref{Fig:Min_sec_cha_M} clearly show the vulnerability of CF-mMIMO to the passive eavesdropping when a carefully designed power control is not employed.

In Figs. \ref{Fig:Min_sec_cha_J} and \ref{Fig:Min_sec_cha_K}, we repeat the experiment in Fig. \ref{Fig:Min_sec_cha_M} with different values of $J$ and $K$. As $J$ increases in Fig. \ref{Fig:Min_sec_cha_J}, the secrecy performance gets deteriorated, i.e., $F(r_s; J, K) \le F(r_s; J', K)$ for $J \le J'$ where $F(r_s; J, K)$ is the CDF at the minimum secrecy rate $r_s$ with $J$ eavesdroppers and $K$ users for a scheme since the maximum leakage rate among Eves increases by the principle of order statistics. Meanwhile, as $K$ increases in Fig. \ref{Fig:Min_sec_cha_K}, both the existing scheme and the proposed scheme without AN have better performances, i.e., $F(r_s; J, K) \ge F(r_s; J, K')$ for $K\le K'$. This happens due to the fact that as $K$ increases, the interference caused by the data signals for other users (which works similarly to the AN) grows, which significantly reduces the leakage rate. It should also be mentioned that as $K$ increases, the data rates of users in all the schemes decrease since the APs with limited transmit power should serve more users simultaneously. However, the reduced leakage rate is more influential than the decrease of user rates. Thus, the existing scheme and the proposed scheme without AN end up with improved performances.

On the contrary, the performance of the proposed AN-aided scheme gets lowered, which is due to the fact that the role of AN is in part replaced with the growing interference caused by the increasing number of users. While the AN power can be saved due to the interference, the decrease of transmit power per user with the increasing number of users is considerably larger, which makes the performance of the proposed AN-aided scheme reduced. As the number of users grows to the extent that the role of AN is completely replaced with the user interference, it is expected that the performances of the proposed scheme with/without AN meet. The proposed AN-aided scheme, nevertheless, always has better secrecy performance than those of the competing schemes. The results in Fig. \ref{Fig:Min_sec_cha_K} elucidate that the role of AN signal for the secrecy performance gets more prominent as the number of users in CF-mMIMO decreases.

\section{Conclusion}
In this work, we investigated secure communications in CF-mMIMO under passive eavesdropping. For maximizing the minimum secrecy rate, we proposed the AN-aided secure power control scheme. By conducting performance comparisons, it was shown that the proposed AN-aided secure power control scheme significantly improves secrecy performance. 

\appendices
\section{} \label{Sec:App_rate}
We first derive the data rate of the $k$th user from the received signal in \eqref{Eq:r_k}. 
By applying the UatF lower bound, the data rate of the $k$th user, $R_k$, can be rewritten as
\begin{align}
	 R_k \ge & \bar{R}_k  = \log_{2} \left( 1 + SINR_k \right)  \nonumber\\
& =\log_{2} \left(1 +  \frac{|\mathbb{E}[f_{kk}]|^2}{\mathbb{V}(f_{kk}) + \sum_{\substack{k'=1 \\ k' \neq k}}^{K} \mathbb{E}[| f_{kk'}|^2] +  \mathbb{E}[| z_{k}|^2] + 1} \right), \nonumber 
\end{align}
where $\mathbb{V}(f_{kk})=\mathbb{E}[|f_{kk}-\mathbb{E}[f_{kk}]|]^2$.
By the definition of $\hat{g}_{mk}$ in \eqref{Eq:g_mk}, $\mathbb{E}[f_{kk}]$ can be readily derived as 
$\sum_{m=1}^M \sqrt{p_{mk}}\gamma_{mk}$ where $\gamma_{mk} = \mathbb{E}[g_{mk} \hat{g}_{mk}^* ]=\sqrt{\tau_p p_p} \beta_{mk}c_{mk}$. 
The derivations of $\mathbb{E}[|f_{kk}-\mathbb{E}[f_{kk}]|]^2$ and $\mathbb{E}[| f_{kk'}|^2]$ can be found in \cite[Appendix A]{Ngo17Cell-free}.
In addition, $\mathbb{E}[|z_{k}|^2]$  can be obtained as
\begin{align}
\mathbb{E}[|z_{k}|^2]&=\mathbb{E}\left[\left|\sum_{m=1}^M \sqrt{p_{mv}} g_{mk} v_m\right|^2\right] \nonumber \\
&=\sum_{m=1}^M p_{mv} \mathbb{E}\left[  \left| g_{mk} \right|^2 \left| v_m \right|^2 \right] = \sum_{m=1}^M p_{mv} \beta_{mk}. \label{Eq:z_k_derivation}
\end{align}

With the assumption that all Eves have perfect knowledge of the channels, the leakage rate of the $k$th user at the $j$th Eve, $R_{jk}^e$, is given by
\begin{align}
	R_{jk}^e   \le \bar{R}_{jk}^e & = \mathbb{E} \left[ \log_{2} \left( 1 + SINR_{jk}^e \right) \right] \nonumber  \\
	& = \mathbb{E} \left[ \log_{2} \left(1 +  \frac{|f_{jk}^e|^2}{\sum_{\substack{k'=1 \\ k' \neq k}}^{K} |f_{jk'}^e|^2 + |z_j^e|^2+1}\right) \right] \nonumber \\
	& \stackrel{(a)}{\approx}  \log_{2} \left(1 +  \frac{\mathbb{E} \left[|f_{jk}^e|^2\right]}{\sum_{\substack{k'=1 \\ k' \neq k}}^{K} \mathbb{E} \left[ |f_{jk'}^e|^2\right] + \mathbb{E} \left[ |z_j^e|^2 \right]+1}\right), \nonumber
\end{align}
where $(a)$ is from the approximation in \cite[Lemma 1]{Zhang14Power}. 
Similar to the derivation of $\mathbb{E}[|z_{k}|^2]$ in \eqref{Eq:z_k_derivation}, $\mathbb{E}[ |z_j^e|^2 ]$ can be rewritten as $\sum_{m=1}^M p_{mv} \beta_{mj}^e$.
Finally, $\mathbb{E}[|f_{jk}^e|^2]$ is given by
\begin{align}
	\mathbb{E} \left[|f_{jk}^e|^2\right] & = \mathbb{E} \left[ \left| \sum_{m=1}^M \sqrt{p_{mk}}\hat{g}^*_{mk} g_{mj}^e \right|^2  \right] \nonumber \\
	&= \sum_{m=1}^M p_{mk} \mathbb{E}\left[  \left| \hat{g}_{mk}^* \right|^2 \left| g_{mj}^e \right|^2 \right] = \sum_{m=1}^M p_{mk} \gamma_{mk} \beta_{mj}^e.  \nonumber
\end{align}

\section{} \label{Sec:App_bound}
For the linear approximation of \eqref{Eq:opt_user_rate_const_lower}, we utilize the convexity of composite functions and the first-order multivariate Taylor series.
Firstly, $-\log_2(1-t)$ is convex and non-decreasing when $0\le t<1$, and $x^2/z$ is also convex when $x \ge 0$ and $z \ge 0 $. Since $-\log_2(1-x^2/z)$ is convex when $ 0 \le x^2 \le z$,
we can derive the lower bound on $-\log_2(1-x^2/z)$ as follows: 
\begin{align}
	-\log_2 & \left(1-\frac{x^2}{z}\right)   \ge -\log_2  \left(1-\frac{x^2_0}{z_0}\right) \nonumber \\
& + \frac{2x_0}{\ln2 \left(z_0 - x^2_0\right)}\left(x-x_0\right)-\frac{x^2_0}{ \ln2 \left( z^2_0 - z_0 x^2_0 \right)}\left(z-z_0\right). \nonumber
\end{align}
By letting $y=z-x^2$, we can obtain the following inequality after some algebraic manipulations.
\begin{align}
	\log_2  \left(1+\frac{x^2}{y}\right) & \ge \log_2  \left(1+\frac{x^2_0}{y_0}\right) \nonumber \\
	 + \frac{1}{\ln2} & \frac{x^2_0}{y_0+x^2_0}\left( \frac{2x-x_0}{x_0} - \frac{y}{y_0} + \frac{(x-x_0)^2}{y_0}  \right) \nonumber \\
	& \ge \log_2  \left(1+\frac{x^2_0}{y_0}\right) + \frac{1}{\ln2}\frac{x^2_0}{y_0+x^2_0}\left( \frac{2x-x_0}{x_0} - \frac{y}{y_0} \right). \nonumber 
\end{align}

Also, by using the first-order multivariate Taylor series, we can obtain the linear approximation of \eqref{Eq:opt_Eve_rate_const_app} as follows:
\begin{align}
	 \log_2  \left(1+\frac{x}{y}\right) & \approx  \log_2  \left(1+\frac{x_0}{y_0}\right) \nonumber \\
	   + \frac{1}{\ln2} & \frac{1}{x_0+y_0} \left(x-x_0\right) - \frac{1}{\ln2}\frac{1}{x_0+y_0}\frac{x_0}{y_0} \left(y-y_0\right) \nonumber \\
	& = \log_2  \left(1+\frac{x_0}{y_0}\right) + \frac{1}{\ln2} \frac{x_0}{x_0+y_0} \left( \frac{x}{x_0}- \frac{y}{y_0} \right). \nonumber
\end{align}

\bibliographystyle{IEEETran}

\end{document}